\documentclass[12pt]{article}
\usepackage{epsfig}
\begin{document}
\begin{titlepage}
\begin{center}
%\hfill{SNUTP-05-011}\\

  \vspace{2cm}

  {\Large \bf X(1835) as  the Lowest Mass Pseudoscalar Glueball
   and Proton Spin Problem}
  \vspace{0.50cm}\\
  Nikolai Kochelev$^{a,b}$\footnote{kochelev@theor.jinr.ru},
  Dong-Pil
Min$^a$ \footnote{dpmin@phya.snu.ac.kr} \vspace{0.50cm}\\
{(a) \it School of Physics and Center for Theoretical Physics,
Seoul National University,
  Seoul 151-747, Korea}\\
\vskip 1ex {(b) \it Bogoliubov Laboratory of Theoretical Physics,
Joint Institute for Nuclear Research, Dubna, Moscow region, 141980
Russia} \vskip 1ex
\end{center}
\vskip 0.5cm \centerline{\bf Abstract} We consider the parity
doublet structure observed in high hadronic excitations   within the
instanton model for the QCD vacuum. In the conventional approach this
doubling phenomenon is treated as a manifestation of the partial restoration of
chiral or $U(1)_A$ symmetry. We demonstrate  that the suppression
of direct instanton contribution to the masses of excited hadrons
leads to the partial $U(1)_A$ symmetry restoration in hadron
spectrum. The origin of
X(1835) resonance observed by BES Collaboration is studied upon the doublet structure. We argue also how
X(1835) be interpreted as the lowest pseudoscalar glueball state,
and derive its coupling constant to proton.
It turns out that this coupling  is large and negative.
Demonstrated is how this large coupling affects the gluonic contribution to the proton spin.

\vspace{1cm}
\end{titlepage}
\setcounter{footnote}{0}
\section{Introduction}

Recently, several attempts have been made to explain the observed
parity doublet structure in excited hadron spectrum \cite{glozman,
shifman}. In fact, Glozman \cite{glozman} suggested, without
giving the underlying mechanism $ a {\ }la $ QCD, that the
appearance of this structure is the consequence of the restoration
of the chiral or $U(1)_A$ symmetry in  highly excited states.

  The existence of instanton, a
strong non-perturbative fluctuation of gluon field, in QCD vacuum
is considered as a primary factor for the chiral
  and $U(1)_A$ symmetries violation in strong interaction (see reviews
\cite{shuryak, diakonov}). Therefore it is natural to assume that
the parity doublet structure in high hadronic excitations stems
from  suppression of the interaction between quarks and gluons
induced by  instanton. A well-known example of such
interaction is the famous t'Hooft quark-quark interaction arising
from the quark zero-modes in instanton field \cite{thooft}.
 The most convenient way to
investigate the properties of hadrons is the consideration  of the
correlator of the currents with definite quantum numbers, for
instance  QCD sum rule method \cite{QCDSR}. Within this approach
the $U(1)_A$-violated t'Hooft interaction  is related to the
so-called direct instanton contribution to the
correlator~\cite{shuryak}.

 In this Letter we show that
for high hadronic excitations   the  direct instanton contribution
to the difference of the correlators of  currents with
the same quantum numbers except parity
 is suppressed and therefore the $U(1)_A$ should be partially restored. As a
result of the restoration, the parity doublet structure in
hadron spectrum is expected.
We can obtain the relationship of residues and masses of parity
partners with the single instanton contribution to correlators.
  As an  important application
 we try to interpret the X(1835) resonance recently observed by BES
Collaboration \cite{BES1, BES2}
 as a glueball.
  The arguments to
consider the parity doublet $[f_0(1710),X(1835)]$ as the lowest
glueball states are given. We present the estimations of X(1835)
coupling to gluons and
 its contribution to the  generalized Goldberger-Treiman relation
 for the
flavor singlet axial coupling of proton. We then show how the so-called
proton spin problem may be resolved.

\section{Parity doubling in hadron spectrum and instantons}

  A very important feature of
the  direct instanton contribution to the correlator of two
hadronic currents with the  opposite parities but with the same
other quantum numbers, is the flipping of the sign of contribution
with changing of parity
\begin{equation}
i\int d^4xe^{iqx}<0|TJ_{\pm}(x)J_{\pm}(0)|0>=\pm \Pi_{I}(Q^2),
\label{parity}
\end{equation}
where $Q^2=-q^2$. This property is  known for a long time (see for
example Tables in paper \cite{carlitz}) and directly related to
the $U(1)_A$ symmetry violation induced by instantons and to the
(anti-) self-duality of (anti-)instanton field in Euclidean
space-time, $G^{I, \bar I}_{\mu\nu}=\pm\widetilde G^{I, \bar
I}_{\mu\nu} $.
  Therefore, the difference in two
correlators, determining the splitting of mass among a parity
doublet,
\begin{equation}
\Delta\Pi(Q^2)=i\int
d^4xe^{iqx}(<0|TJ_+(x)J_+(0)|0>-<0|TJ_-(x)J_-(0)|0>),
\label{delta}
\end{equation}
can be related to the single-instanton contribution calculated
with some effective instanton density. On the other hand, the
phenomenological representation of Eq. (\ref{delta}) can be
determined by the difference of contributions of parity partners
\begin{equation}
\Delta\Pi(Q^2)=\sum_n\{
\frac{{\lambda_{n+}}^2}{{M_{n+}}^2+Q^2}-\frac{{\lambda_{n-}}^2}{{M_{n-}}^2+Q^2}\},
\label{diff}
\end{equation}
where $\lambda_i$ is the residue, and $M_i$  is the mass of
corresponding resonances.

The instanton field has the strong localization  in space-time
\begin{equation}
{A_\mu}^a(x)=\frac{2\bar\eta_{a\mu\nu}(x-x_0)_\nu\rho^2}{g_s(x-x_0)^2((x-x_0)^2+\rho^2)},
\label{inst}
\end{equation}
where $x_0$ is the center of instanton , $\rho$ is the instanton
size, $\bar\eta_{a\mu\nu}$ is the 't Hooft symbol. Eq.
(\ref{inst}) corresponds to the singular gauge. The localization
property leads to the direct instanton contribution into hadron
correlator which is proportional to the product of McDonald
functions $K_n(z)$, with $z=\rho Q$ \cite{shuryak1}. The McDonald
function has exponential damping factor  $\sim exp(-\rho Q)$ at
large $z$ . The contribution from the given parity doublet becomes
important in the sum in Eq.(\ref{diff}) at  $|Q|\sim M_i $.
Therefore, in high hadronic excitation, due to the exponential
decay of direct instanton contribution to  $\Delta\Pi(Q^2)$, we
expect the decrease of splitting in a parity doublet. It is
evident now that the observed degeneration  between excited hadron
states of opposite parities can be understood as an effect of the
partial restoration of the $U(1)_A$ symmetry arising from the
suppression of single-instanton contribution. It should be
emphasized that the chiral symmetry
 restoration does not result from the same suppression mechanism because
  the chiral symmetry violation is intimately related to
   the delocalization of quarks among many instantons~\cite{diakonov}.

 \section{X(1835) as a lowest mass pseudoscalar glueball}

The identification of the glueball states in the hadron spectrum
is one of the most exciting topics of hadron spectroscopy
\cite{reviews, bugg,anisovich, vento}. In the scalar sector of
$J^{PC}=0^{++}$ there are three candidates for the most low-lying
mass glueball states as $f_0(1370)$, $f_0(1500)$ and $f_0(1710)$.
Their masses are  in fair agreement with the prediction of lattice
QCD , $ m=1.4 \sim 1.8$ GeV, \cite{lattice} and also with QCD sum
rules approaches \cite{narison, forkel}. However, for the
pseudoscalar sector of ($J^{PC}=0^{-+}$), the situation is not so
clear. Only $\eta(1440)$ is being discussed as a possible
candidate for the pseudoscalar glueball \cite{minkowski, faddeev}.
But, this identification is doubted by
 large disagreement between its experimental observation and
theoretical predictions of both lattice QCD \cite{lattice} and QCD
sum rules calculation \cite{narison, forkel}, of which values
range between $m=1.86$ and $ 2.7$ GeV. Furthermore, the
peculiarities of $\eta(1440)$ production in different reactions
support these doubts on its glueball origin \cite{klempt, close}.

Recently, BES Collaboration observed X(1835) state in the
reactions $J/\Psi\rightarrow \gamma p\bar p$ and
$J/\Psi\rightarrow \gamma \eta^\prime \pi^+\pi^- $ \cite{BES1,
BES2}. The most remarkable property of  this state is its strong
coupling with proton-antiproton channel, leading to speculations
about its baryonium origin (see \cite{zhu} and references
therein).

Now let us  apply  the instanton mechanism for parity doubling in
hadron spectrum discussed above to the problem of low mass
glueballs. \footnote{The phenomena of the parity degeneracy in
glueball spectrum is also appeared within model of glueballs as
the twisted closed gluonic flux tubes \cite{faddeev}.} The direct
instanton contribution to the difference of two correlators of
glueball currents with opposite parities  is given by
\cite{novikov,forkel}
\begin{eqnarray}
\Delta\Pi(Q^2)_G&=&i\int
d^4xe^{iqx}(<0|TO_S(x)O_S(0)|0>-<0|T|O_P(x)O_P(0)|0>) \nonumber\\
&=&2^6\pi^2\int_0^{\rho_{cut}} d\rho n(\rho)(\rho Q)^4{K_2}^2(\rho
Q), \label{direct}
\end{eqnarray}
where the glueball currents for scalar and pseudoscalar states are
the following
\begin{eqnarray}
O_S(x)&=&\alpha_sG_{\mu\nu}^a(x)G_{\mu\nu}^a(x),\label{pseudo}\\
O_P(x)&=&\alpha_sG_{\mu\nu}^a(x)\widetilde
G_{\mu\nu}^a(x),\label{scalar}
\end{eqnarray}
and $\widetilde
G_{\mu\nu}^a(x)=1/2\epsilon_{\mu\nu\alpha\beta}G_{\alpha\beta}^a(x)$.
In Eq. ({\ref{direct}}), $\rho_{cut}$ is the  cutoff of instanton
size, $\rho_{cut}\approx 1/\mu_r$, where $\mu_r$ is the
normalization  scale  (see discussion in \cite{forkel}).

  Within the instanton liquid model of QCD
vacuum the instanton density $n(\rho)$ in Eq. \ref{direct} can be
approximated rather well by the Gaussian-tail distribution ( see
\cite{shuryak} and discussion around Eq. (85) in \cite{forkel})
\begin{equation}
n(\rho)=\frac{2^{18}n_{eff}\rho^4}{3^6\pi^3\rho_c^5}exp({-\frac{2^6\rho^2}{3^2\pi\rho_c^2}})
\label{density},
\end{equation}
with $n_{eff}\approx 0.5 fm^{-4}$ and $\rho_c\approx 0.33 fm$. The
shape of Eq.(\ref{density}) has correct behavior as being $\propto
\rho^{b_0-5}$ at small $\rho$ for number of flavors $N_f=3$ (
$b_0=11N_c/3-2N_f/3$) given by instanton perturbative theory, and
at large $\rho$ as being $\propto exp(-a^2\rho^2)$ which is
supported by the lattice calculations (see discussion in
\cite{schrempp}).

We may match the glueball contribution of Eq.(\ref{direct}) with
the phenomenological Eq.(\ref{diff}) by noting the most important
contribution comes from the low-lying mass states. Then, all
states can be identified with quantum numbers $I=0,
J^{PC}=0^{\pm,+}$. For the negative parity states we have
evidently the following candidates as $\eta(550)$,
$\eta^\prime(958)$, $\eta(1295)$, and $\eta(1440)$.~\footnote{The
Particle Date Group
  gives now two resonances $\eta(1405)$ and $\eta(1475)$, instead
  of a single resonance
  $\eta(1440)$ \cite{PDG}. However, this splitting can be easily
  explained by a node in the wave function of $\eta(1440)$,
  interpreted as a first radial excitation of the $q\bar q$
  system \cite{klempt}.}
  We now  assume that the new resonance X(1835) is the lowest  mass
pseudoscalar glueball state and will be added to this list. For
the positive parity states the candidates are $f_0(600)/\sigma$,
$f_0(980)$, $f_0(1370)$, $f_0(1500)$, and $f_0(1710)$. Now, by
looking at the values of their masses one can form  the   parity
doublets :  $ [\eta(550), f_0(600)/\sigma] $, $[\eta^\prime(958),
f_0(980)]$,
  $[\eta(1295), f_0(1370)]$, $[\eta(1440), f_0(1500)]$, $[X(1835),
f_0(1710)]$. \footnote{Recently BES Collaboration has observed
additional resonance $f_0(1790)$ in the decay
   $J/\Psi\rightarrow\phi\pi^+\pi^-$ \cite{BES3}. This state has
similar properties with
   $f_0(1370)$, which has observed in the same reaction.
   Therefore, one can treat $f_0(1790)$ as the next radial
   excitation of $f_0(1370)$. Furthermore, this state is a good
candidate for the  parity partner
    of $\eta(1760)$ \cite{PDG}.
   However, it is necessary to confirm the existence of both of these
states.}
   We  consider only
chiral limit $m_u=m_d=m_s=0$ and therefore neglect all possible
effects related to the mixing between $SU(3)_f$ octet and singlet
states. Thus, the contribution to Eq.(\ref{diff}), originated from
member of $SU(3)_f$ octet, $\eta(550)$, and, correspondingly, its
parity partner $f_0(600)/\sigma$,  should be excluded. It is well
known that due to axial anomaly $\eta^\prime(958)$ has strong
coupling to gluons (see review \cite{shifman2}):
\begin{equation}
N_f<0|\frac{\alpha_s}{4\pi}G_{\mu\nu}^a\widetilde{G_{\mu\nu}^a}|\eta^\prime>
  =F_{\eta^\prime}m_{\eta^\prime}^2,
  \label{residueeta}
  \end{equation}
where $F_{\eta^\prime}\approx\sqrt{3}f_\pi,$  the $f_\pi= 132$ MeV
is the pion decay constant. Therefore, there is the contribution
to Eq. (\ref{diff}) coming from the doublet $[\eta^\prime(958),
f_0(980)]$. Usually, $\eta(1295)$ and $\eta(1440)$ are considered
as good candidates for $2S$ radial excited states of $\eta$ and
$\eta^\prime$. We accept this treatment and  neglect contributions
from  doublets $[\eta(1295), f_0(1370)]$ and $[\eta(1440),
f_0(1500)]$ to Eq. (\ref{diff}) which must be smaller than the
contribution from the doublet $[\eta^\prime(958),f_0(980)]$. Our
numerical fit shows that even the  contribution of the ground
$q\bar q $ state
  mesons,
   $\eta^\prime(958)$ and $f_0(960) $, to Eq.\ref{diff} is very small.
Therefore, one can neglect all
   contributions of
    their radial excitations.

Let us discuss now the candidate for glueball parity doublet
  $[X(1835), f_0(1710)]$. Both of them have been
observed in radiative $J/\Psi$ decay, which is considered as one
of the important properties of glueball. Furthermore,
   in a recent
paper \cite{chanowitz} it was argued that the larger observed
coupling of scalar glueball  $f_0(1710)$  to $K\bar K$ state than
to two pion state could be easily explained by using the chirality
argument. The $X(1835)$ was observed in two channels, $p\bar p$
and $\eta^\prime\pi^+\pi^-$, but was not in $\pi^0\pi^+\pi^-$
channel. The observation of $X(1835)$ in the channel with
$\eta^\prime$ suggests strongly its glueball origin stemming from
the large coupling of $\eta^\prime$ to gluons. Furthermore, the
strong coupling of this state to proton-antiproton state  can  be
related to the large contribution of gluon axial anomaly to proton
spin as we will show later. The non-observation of decay of this
state to $\pi^0\pi^+\pi^-$  can be easily explained by very small
mixing, being $\propto (m_u-m_d)/M_G$, of $I=1$ $\pi^0 $ state
with $I=0$ glueball. We should also mention that the mass value of
$f_0(1710)$ for the scalar glueball is supported by the quenched
lattice  calculations \cite{lattice}, while the mass value of
$X(1835)$ for the pseudoscalar glueball state is smaller than the
quenched lattice prediction, $M_{0^-}\approx 2.5$ GeV
\cite{lattice}. However, such a difference might be related to the
neglect of the correlations between instantons in the quenched
approximation. These correlations come from the virtual $q\bar q$
pair exchanges  between instantons. One should keep in mind that
such correlations lead to the so-called topological charge
screening effect, which is specially important for flavor singlet
pseudoscalar channel (see discussion in \cite{shuryak} and
\cite{forkel}). It would not be worthless to point out that the
mass of X(1835) is quite close to the lowest value predicted by
QCD sum rules for pseudoscalar glueball $M_P=1.86$ GeV
\cite{narison} and $M_P=2$ GeV \cite{forkel}.

Now we are in position to  estimate the coupling of $X(1835)$ with
gluons. By assuming that residues of parity partners are
equal~\footnote{The attempt to fit the difference of correlators
Eq. (\ref{direct}) with the different values of residues in Eq.
(\ref{diff}) has lead to the worse result.} and using Eq.
(\ref{residueeta})
\begin{equation}
  \lambda_{f_0(980)}=
\lambda_{\eta^\prime}=<0|\alpha_sG_{\mu\nu}^a\widetilde{G}_{\mu\nu}^a|\eta^\prime>\approx
0.88 GeV^3,
\end{equation}
\label{lambda1} the  value for the residue of $X(1835)$
\begin{equation}
\lambda_{f_0(1710)}=
  \lambda_{X}=<0|\alpha_sG_{\mu\nu}^a\widetilde{G}_{\mu\nu}^a|X>,
  \label{lambda2}
  \end{equation}
  is
obtained by fitting the contribution of direct instantons Eq.
(\ref{direct}) with a natural normalization scale $\mu_r\approx
M_{\eta^\prime}$ by formula
\begin{equation}
\Delta \Pi (Q^2)=
\lambda_{\eta^\prime}^2(\frac{1}{M_{f_0(980)}^2+Q^2}-
\frac{1}{M_{\eta^\prime}^2+Q^2})+
\lambda_{X}^2(\frac{1}{M_{f_0(1710)}^2+Q^2}-
\frac{1}{M_{X(1835)}^2+Q^2}) \label{fit}
\end{equation}
  in interval $Q>1 GeV$.
The final result is
\begin{equation}
\lambda_X=\lambda_{f_0(1710)}=2.95 GeV^3. \label{gcoupling}
\end{equation}
The  value Eq.(\ref{gcoupling}) can be compared with results of
recent QCD sum rules analysis for pseudoscalar  $ \lambda_P\approx
2.9$ GeV $^3$ and scalar $\lambda_S\approx 1.64$ GeV $^3$
glueballs \cite{forkel} (see also \cite{narison}).
\begin{figure}[htb]
\centerline{\epsfig{file=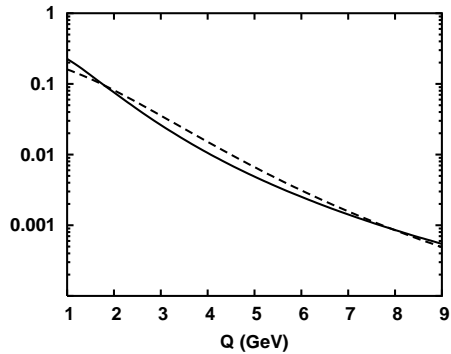,width=6cm,angle=0}}\
\caption{The result of  fitting the phenomenological part of
difference between two correlators of gluon currents, Eq.
(\ref{fit}), (dashed line) in comparison with the direct instanton
contribution, Eq. (\ref{direct}), (solid line).} \label{fig1}
\end{figure}
  The comparison of theoretical Eq.(\ref{direct}) and
phenomenological Eq.(\ref{fit}) results  is presented in Fig.1. It
is evident that the our phenomenological model for the difference
of two gluon correlators, based on interpretation of $X(1835)$ as
a lowest glueball state, gives very good description of direct
instanton contribution. We should also mention, that due to large
difference in the value of residues between glueball $[X(1835),
f_0(1710)] $ and quark $[\eta^\prime(958),f_0(980)]$
  doublets, the
contribution of later can be safely neglected. Thus, one can
conclude that our hypothesis on glueball origin of $X(1835)$ is
not in the contradiction with the modern knowledge of the behavior
of gluon current correlators, and the coupling of this state to
gluons is very large\footnote{The possibility of strong coupling
X(1835) to gluons was pointed out by Rosner in \cite{rosner2}}.

\section{ Proton spin problem and  glueball $X(1835)$}

  During the last two decades there have been many attempts, both
theoretical and experimental (see reviews \cite{jaffe, anselmino,
dorkoch}), to understand how the spin is distributed among the
different components of the proton. This investigation is named
generically as  the proton spin problem. Since no fully
satisfactory understanding albeit many conjectures has been found
to this problem there is no unique answer to the question: where
lies the proton spin? In this section a new way of the solution to
that problem  based on the existence of light pseudoscalar
glueball $X(1835)$ is suggested.

Present wisdom tells that the small observed value of the flavor
singlet axial  coupling of nucleon $g_A^0$, is followed from the
non-conservation of the flavor singlet axial-vector current in
chiral limit
\begin{equation}
\partial_\mu J_{\mu 5}^0(x)= 2N_f\frac{\alpha_s}{8\pi}G_{\mu\nu}^a
\widetilde{G}_{\mu\nu}^a. \label{anom}
  \end{equation}
By taking the matrix element of Eq. (\ref{anom}) between proton
states, one can find the relation \cite{veneziano1,zakharov}
  \begin{equation}
  g_{A}^0i\bar\Psi_P\gamma_5\Psi_P =
\frac{N_f}{M_P}<P|\frac{\alpha_s}{8\pi}G_{\mu\nu}^a
\widetilde{G}_{\mu\nu}^a|P>,
  \label{glu}
   \end{equation}
where $M_P$ is the proton mass. The matrix element of gluon
operator in Eq.(\ref{glu}) can be rewritten as a sum over all
possible intermediate pseudoscalar states $G$ connected with
gluons by
\begin{eqnarray}
  <P|\alpha_sG_{\mu\nu}^a
\widetilde{G}_{\mu\nu}^a|P>=\sum_{G,{\ } k^2 \rightarrow
0}\frac{<0|\alpha_sG_{\mu\nu}^a
\widetilde{G}_{\mu\nu}^a|G><GP|P>}{k^2-M_{G}^2},
  \label{matrix}
   \end{eqnarray}
where the coupling $<GP|P>$ is
\begin{equation}
<GP|P>=-ig_{GPP}\bar\Psi_P\gamma_5\Psi_P.  \label{inter1}
\end{equation}
  By using
Eqs. (\ref{lambda1}, \ref{lambda2}, \ref{matrix}), we obtain
\begin{equation}
g_A^0=\frac{F_{\eta^\prime}g_{\eta^\prime
PP}}{2M_P}+\frac{F_Xg_{XPP}}{2 M_P}. \label{ga}
\end{equation}
where
\begin{equation}
  F_X=\frac{N_f\lambda_X}{4\pi M_X^2}.
\label{def}
\end{equation}
   We will call
Eq. (\ref{ga}) as a generalized Goldberger-Treiman relation for
the flavor singlet axial coupling in comparison with the
well-known Goldberger-Treiman relation for the flavor non-singlet
axial vector coupling in the chiral limit
\begin{equation}
g_A^3=\frac{F_\pi g_{\pi NN}}{2M_P}, {\ \ } g_A^8=\frac{F_\eta
g_{\eta NN}}{2M_P} \ , \label{usual}
\end{equation}
where $F_\pi=\sqrt{2}f_\pi$ and $F_\eta\approx \sqrt{6}f_\pi$.
  The first term in
Eq. (\ref{ga})  can be theated as the valence quark contribution
to the proton spin. For $ SU(6)$ value of $g_{\eta^\prime NN}=6.5$
(see \cite{anselmino}) we have for the valence part
\begin{equation}
g_A^{v}=0.79, \label{valence}
\end{equation}
which lies between the values given by the non-relativistic quark
model, $g_A^v=1$ and by the MIT bag model $g_A^v=0.65$. The second
term in Eq. (\ref{ga}) comes from $X(1835)$ glueball and therefore
its  interpretation   as a gluon contribution to proton spin is
very plausible.
  The gluon contribution is determined by the
value of residue Eq. (\ref{gcoupling}), which has been estimated
above, and by the value of coupling $g_{XPP}$. This coupling was
estimated in a recent paper \cite{zhu} from experimental branching
ratio of production $X(1835)$ to $p\bar p$ in radiative decay of
$J/\Psi$
\begin{equation}
\frac{{g_{XPP}}^2}{4\pi}\approx 1 , \label{bes1}
\end{equation}
i.e.  $|g_{XPP}|\approx 3.5$. We should stress that the  sign of
the coupling cannot be fixed via the branching ratio. The value of
$g_{XPP}$ is very large,  approximately one half of $\eta^\prime$
coupling with nucleon. The arguments in favor of negative sign of
this coupling constant will be given below. It follows that
\begin{equation}
g_{XPP}\approx -3.5. \label{bes}
\end{equation}
By using Eqs. (\ref{gcoupling}, \ref{ga}, \ref{bes}) we obtain for
the gluon contribution to $g_A^0$
\begin{equation}
g_A^{glue}\approx -0.39. \label{gaglue}
\end{equation}
The total value of the  flavor singlet axial coupling is
\begin{equation}
g_A^0\approx 0.4. \label{theory}
\end{equation}
Therefore, we have obtained very large decreasing of $g_A^0$,
arising from the contribution of glueball $X(1835)$ to proton
matrix element of pseudoscalar gluon density. The value of
$g_A^0$, Eq. (\ref{theory}),  is in agreement with the present
experimental data $g_A^0\approx 0.3\sim 0.4 $ (see \cite{sidorov},
\cite{forte} and references therein).

To estimate  the coupling of $X(1835)$ to proton within
non-perturbative QCD  we assume that  the coupling is determined
by  interaction of its two valence gluons with proton quarks
through the instanton. The  two-gluon interaction with quark
induced by instanton follows from the generalized t'Hooft
Lagrangian obtained from quark and gluon fields in instanton
background \cite{thooft,ABC,shuryak,diakonov}:
\begin{eqnarray}
{\cal L}_{eff}&=&\int dUd\rho n(\rho)\prod_q -F_q(k_q\rho,k_g\rho)
\frac{2\pi^2\rho^3}{m_q^*\rho}
  \bar q_R(1+\frac{i}{4}U_{ab} \tau^a
\bar\eta_{b\mu\nu}\sigma_{\mu\nu})q_L
\nonumber\\
&\cdot&
e^{-\frac{2\pi^2}{g}\rho^2U_{cd}\bar\eta_{d\alpha\beta}G^c_{\alpha\beta}}+(R\leftrightarrow
L), \label{lag}
\end{eqnarray}
where $m_q^*$ is the effective quark mass in the instanton vacuum,
$U$ is
  the orientation matrix of the instanton in $SU(3)_c$. Notice that
the form factor $F_q$ takes into account  the off-shell of quarks
and gluons interacting to the instanton, and $k_{q}$, $k_g$ are
virtualities of quarks and gluons, respectively. The Lagrangian
Eq. (\ref{lag}) leads to the following two-gluon vertex coupling
with flavor singlet pseudoscalar current $J_{5}^0=\bar u\gamma_5
u+\bar d\gamma_5 d+\bar s\gamma_5 s$
\begin{equation}
{\cal L}_{ggq}=-iF(Q)\frac{n_{eff}\pi^3\rho_c^4}{4<0|\bar q
q|0>\alpha_s^2}\alpha_s G_{\mu\nu}^a \widetilde{G}_{\mu\nu}^a
J_{5}^0, \label{gqq}
\end{equation}
where $F(Q)$ is a form factor. In vacuum dominance
approximation\footnote{The similar approximation was used in
papers \cite{kv} and \cite{schafer} for the estimation of the
instanton contribution to weak decay $K\rightarrow\pi\pi$ and
strong decays of charmonium and glueballs.} the effective
$X(1835)$ interaction is given by
\begin{equation}
{\cal L}_{Xqq}=-ig_{Xqq}XJ_{5}^0 \label{Xqq1},
\end{equation}
where
\begin{equation}
g_{Xqq}\approx \frac{\pi^3n_{eff}\rho_c^4\lambda_X}{4<0|\bar
qq|0>\alpha_s^2}F_X(m_X),
  \label{Xqq}
  \end{equation}
and we have elaborated the simplest version of Shuryak's instanton
liquid model \cite{shuryak1}:
\begin{equation}
n(\rho)=n_{eff}\delta(\rho-\rho_c), {\ }
m_q^*=-\frac{2}{3}\pi^2\rho_c^2<0|\bar q q|0>. \label{shurmod}
\end{equation}

The effective $\eta^\prime$-quark Lagrangian
\begin{equation}
{\cal L}_{\eta^\prime qq}=-ig_{\eta^\prime qq}\eta^\prime J_5^0
\label{eta}
\end{equation}
   follows from the four-quark
t'Hooft interaction induced by instantons \cite{thooft}. Within
the same approximation as above the coupling is derived as follows
\begin{equation}
g_{\eta^\prime
qq}\approx\frac{2n_{eff}{\lambda_{\eta^\prime}}^q}{<0|\bar q
q|0>^2}F_{\eta^\prime}(m_{\eta^\prime})
  \label{etaqq}
  \end{equation}
where ${\lambda_{\eta^\prime}}^q=<0|iJ_5^0|\eta^\prime>/3$, and
$F_{\eta^\prime} (m_{\eta^\prime}) $ is the instanton induced form
factor. With the values ${\lambda_{\eta^\prime}}^q\approx 0.16$
GeV$^2$ \cite{diakeides}, $<0|\bar qq|0>=-(250 MeV)^3 $, and
$\alpha_s\approx 0.5$ \cite{diakonov}, we obtain the estimate
\begin{equation}
g_{Xqq}\approx -36.1\cdot F_X(m_X), {\ } {\ }  g_{\eta^\prime
qq}\approx 1.0\cdot F_{\eta^\prime}(m_{\eta^\prime}).
  \label{numer}
  \end{equation}
So that the coupling constant of pseudoscalar glueball is  {\it
negative}. The absolute value of couplings strongly depends on the
form factors in Eq. (\ref{numer}) which are determined by the
complicate dynamics related to   wave functions of $X(1835)$,
$\eta^\prime$ and instanton form factor. The ratio of couplings of
$X(1835)$ and $\eta^\prime$ can be estimated roughly as
\begin{equation}
\frac{g_{XPP}}{g_{\eta^\prime PP}}\approx
\frac{g_{Xqq}}{g_{\eta^\prime qq}}\approx -36\cdot
exp(-2\rho_c(m_X-m_{\eta^\prime}))\approx -2 .
  \label{num2}
\end{equation}
Therefore, the glueball coupling with proton is large and
negative. This negative sign is  related directly to the negative
sign of quark condensate and  sign difference of effective t'Hooft
interaction for even and odd number of quark legs incoming to the
instanton.

  We should emphasize that at present the both,
$\eta^\prime$ and $X$ couplings with proton, are not well known.
Thus, the  old phase shift analysis of elastic nucleon-nucleon
scattering within  one boson exchange (OBE) model \cite{dum} gives
a large value for $g_{\eta^\prime NN}\approx 7.3$, which is close
to  $SU(6)$ value. Whereas the recent analysis of data on the
reactions $\gamma p\rightarrow p\eta^\prime$ and $pp\rightarrow
pp\eta^\prime $  within relativistic OBE model gives the estimate
$g_{\eta^\prime NN}\leq 3$ \cite{nakayama}. However, it is clear
that for a  careful extraction of $\eta^\prime$ coupling one has
to consider  the additional  glueball contribution to the hadronic
reactions. It should be mentioned  that  rough estimation of the
glueball coupling constant to proton Eq. (\ref{bes}) given in
\cite{zhu} is based on  limited data of the BES Collaboration on
the reaction $J/\Psi\rightarrow \gamma p\bar p$ \cite{BES1}.
Additional experimental and theoretical studies are necessary to
fix  that coupling.

Let us discuss some possible reactions where $X(1835)$ can give
significant  contribution. Our identification of $X(1835)$ as the
lowest pseudoscalar glueball state leads to the expectation that
its coupling to two {\it real} photons should be small or even
vanishing. This expectation is based also on the decoupling of
gluon axial anomaly from two real photons. Thus, the photon
contribution to the axial charge vanishes  on-shell, and the first
moment of spin-dependent structure function of photon $g_1^\gamma
$ becomes zero as well \cite{veneziano3,brodsky}\footnote{For the
real incident photons this fact was known for a long time
\cite{gerasimov}. We are grateful to Sergo Gerasimov for
discussion of this problem.} . However, $X(1835)$ should couple to
heavy vector currents, i.e. to non-zero virtual photons, gluons
and massive  vector mesons. This property  is also related to
peculiarities of axial anomaly (see \cite{anselmino}). Therefore,
one can expect a large contribution of $X(1835)$ in vector meson
photo- and electro-production, in radiative decays of heavy
quarkoniums and in $\gamma^*\gamma$, $\gamma^*\gamma^*$
collisions. In particular, the contribution of $X(1825)$ might be
behind of the difference in cross sections of $\rho^+\rho^-$ and
$\rho^0\rho^0$ production in the reaction
$\gamma\gamma^*\rightarrow\rho\rho$, recently observed by L3
Collaboration \cite{L3}.  Furthermore, the observation of
$X(1835)$
  in the radiative decays $J/\Psi\rightarrow\gamma
(\gamma\rho,\gamma\Phi),\gamma(\rho\rho) $ at BES would give very
useful information on coupling of $X(1835)$ to vector currents.
The investigation of direct photo- and electro-production of
$X(1835)$ at upgraded CEBAF and SPring8 and at COMPASS, together
with the study of its  contribution to the cross-sections of
various reactions at large momentum transfer, is certainly
welcome.

\section{Conclusion}
A new mechanism of  parity doubling in high hadronic excitations
is suggested. We have shown that the direct instanton
contribution, responsible for the mass splitting between states
with opposite parities and the same other quantum numbers, is
suppressed for massive hadrons. Based on the instanton mechanism
of partial $U(1)_A$ symmetry restoration, we consider the parity
doublet $[X(1835), f_0(1710)]$ as the lowest mass glueball states
and suggested an explanation of recent results of BES
Collaboration \cite{BES1}, \cite{BES2}  which observed a resonance
X(1835) with very unusual properties.
  It is shown  that the large coupling
of $X(1835)$ with gluons and proton allows us to give a new
explanation of the so-called 'spin crisis'. The estimation of
glueball coupling to quark, based on effective instanton induced
interaction,  provides the  negative sign of the glueball coupling
to proton state. This negative sign is the fundamental reason for
the observed  smallness of the  flavor singlet axial coupling
constant of nucleon. The relation of the present consideration  to
the approach
 to the proton spin problem based on investigation of QCD
topological susceptibility \cite{narison3,ioffe,dorokhov} will be
the subject of forthcoming publication \cite{kmin1}. It would be
also interesting to study  the connection of $X(1835)$  mechanism
with the gluon contribution to the proton spin obtained within a
chiral bag model \cite{min}, and the relation of this mechanism to
the different approaches based on instanton contribution to proton
spin \cite{kochelev3,schaferspin,shuryakforte}.
\section{Acknowledgments}
We are very grateful to A.E. Dorokhov, S.B. Gerasimov, L.Ya.
Glozman, D.G. Pak, San Fu Tuan and Shi-Lin Zhu
   for useful
discussions. This work was supported by Brain Pool program of
Korea Research Foundation through KOFST,  grant 042T-1-1,  and in
part by grants of Russian Foundation for Basic Research,
RFBR-03-02-17291 and RFBR-04-02-16445 (NK). NK is very grateful to
the School of Physics, SNU, for their warm hospitality during this
work.

\end{document}